# Simulation and Performance Analysis of MP-OLSR for Mobile Ad hoc Networks

Jiazi Yi, Eddy Cizeron, Salima Hamma, Benoît Parrein
Université de Nantes, Nantes Atlantique Universités
IRCCyN, CNRS UMR 6597, Polytech'Nantes, rue Christian Pauc - BP50609
44306 Nantes cedex 3 – France
{jiazi.yi, eddy.cizeron, salima.hamma, benoit.parrein}@univ-nantes.fr

*Abstract* — **Mobile ad hoc networks (MANETs) consist of a collection of wireless mobile nodes which dynamically exchange data without reliance on a fixed base station or a wired backbone network, which makes routing a crucial issue for the design of a ad hoc networks. In this paper we discussed a hybrid multipath routing protocol named MP-OLSR. It is based on the link state algorithm and employs periodic exchange of messages to maintain topology information of the networks. In the mean time, it updates the routing table in an on-demand scheme and forwards the packets in multiple paths which have been determined at the source. If a link failure is detected, the algorithm recovers the route automatically. Concerning the instability of the wireless networks, the redundancy coding is used to improve the delivery ratio. The simulation in NS2 shows that the new protocol can effectively improve the performance of the networks.**

I. INTRODUCTION

Nowadays, great demands for self-organizing, fast deployable wireless Mobile Ad hoc Networks (MANETs) come along with the advances in wireless portable technologies. Compared with the conventional cellular wireless mobile networks that rely on extensive infrastructure to support mobility, the MANETs do not need base stations and wired infrastructure. This future makes it useful in battlefields, emergency searches and rescue operations where fixed base stations are undesirable or unavailable. For commercial applications such as convention centers, electronic classrooms and conferences, a rapid deployment of all-on-air networks provides users with more flexible and cheaper ways to share information.

However, because of the dynamic feature of the Ad hoc networks and the instability of the wireless medium, the routing protocols used in the traditional wired networks seems not be able to be applied in Ad hoc networks. So there are a lot of unipath routing protocols being proposed for Ad hoc networks.

In recent years, more and more multipath routing protocols are also proposed. These protocols consist of finding multiple routes between a source and destination node. These multiple paths between source and destination node pairs can be used to compensate for the dynamic and unpredictable nature of ad hoc networks. The multipath routing could offer several benefits: load balancing, fault-tolerance, higher aggregate bandwidth, lower end-to-end delay, effectively alleviate congestion and bottlenecks [1] and security.

Optimized Link State Routing protocol [2] [3] is a proactive unipath routing protocol specially designed for Ad hoc networks. As a table-driven protocol, it changes control messages periodically to enable the individual nodes be aware of the topology of the whole network. It inherits the stability of the link state algorithm and minimizes the flooding of the control traffic by using only certain selected nodes, called *MPR* (Multi Point Relays).

In this study, we discuses a new multipath routing protocol called MP-OLSR based on OLSR to provide fault-tolerance, higher aggregate bandwidth and load balancing. It exchanges control messages periodically as OLSR to get the topology information of the whole networks. Based on this topology information, our *Multipath Dijkstra* algorithm is used to obtain the multiple paths for the routing. With the algorithm, we can get node-disjoint routes or path-disjoint routes as necessary by adjusting distinct cost functions. In the network, the packets are forwarded from the source to the destination by employing a semi-source routing mechanism (source routing with route recovery). In addition, to meet the need for the reliable transmission, multiple description coding strategy is used in the data transmission. Thus the contribution of the paper is double. First, we propose a multipath routing algorithm based on the Dijkstra algorithm that allows different multipath approaches. Routing recovery from intermediate nodes is also included.

Manuscript received September 19, 2007. This work is supported by the French program RNRT (Réseau National de Recherche en Télécommunications) under the project SEREADMO (http://www.sereadmo. org).

Second, the multiple routes are exploited via an original multiple description coding based on a discrete Radon transform.

The remainder of the paper is organized as follows. Section II gives a brief review of several existing routing algorithms. The specification of the protocol is presented in Section III in detail. In Section IV we introduce the application of Multiple Description Coding in the MP-OLSR. The simulation model and performance results are demonstrated in Section V, and we conclude the paper in Section VI.

## II. ROUTING ALGORITHMS FOR MANET

### A. Proactive Routing and Reactive Routing

Proactive Routing and Reactive Routing are two main kinds of routing protocol for Ad hoc networks [5].

For Proactive Routing, also called table driven routing, each node maintains a routing table containing routes to all nodes in the network. Nodes must periodically exchange messages with routing information to keep routing tables up-to-date. The routing table is calculated before needed. So it has minimal latency but also has high control overhead. The OLSR protocol mentioned above is a typical proactive routing protocol.

For Reactive Routing, also called on-demand routing, a node only tries to find a route when necessary. However, because sometimes the route could not be get immediately, the network using reactive routing usually has longer delay. The DSR (Dynamic Source Routing) [4] is a well-known reactive protocol. In DSR, the source explicitly lists this route in the packet's header, identifying each forwarding "hop" by the address of the next node to which to transmit the packet on its way to the destination host.

Our approach is to get the topology information proactively and compute the routes on-demand.

### B. Multipath Routing

In the literature, multipath routing protocols are often used for backup routes. Otherwise, if the goal is the repartition of information, the implementation is generally based on pure source routing. In [6], the author proposes a scheme called AODV-BR to improve existing on-demand routing protocols by creating a mesh and providing multiple alternate routes. But it just provides backup routes, which means during the transmission, there is still one route is used. In [7], the author proposes an on-demand routing scheme called Split Multi-path Routing (SMR) that establishes and uses multiple routes of maximally disjoint paths. However, as we will see in the following, a pure source routing strategy seems to be not suitable for dense network. The route maintenance by spending Route Error messages will cause longer delay than route recovery.

A source routing multipath OLSR is presented in [8] by using the shortest path algorithm. However, the suppression of nodes in multiple calls of Dijkstra algorithm could not work for sparse networks. Furthermore, strict node-disjoint multiple paths are not suitable for partition or fusion of group of nodes that can imply temporary a single link for connection. And one more time, our work calls into question pure source routing for dense MANET when multiple paths are used.

## III. MP-OLSR SPECIFICATION

The MP-OLSR can be regarded as a hybrid multipath routing protocol. It sends out *HELLO* messages and *TC* messages periodically to be aware of the network topology, just like OLSR. However, MP-OLSR does not always keep a routing table. It only computes the routes when there are data packets need to be sent out.

The core functioning of MP-OLSR has two main parts: *topology sensing* and *routes computation*. The *topology sensing* is to make the nodes get to the topology information of the network, which includes *link sensing, neighbor detection* and *topology discovery*. This part gets benefit from *MPRs* as well as OLSR. The *routes computation* uses the *Multipath Dijkstra Algorithm* to populate the multipath based on the information get from the *topology sensing*. The source route (the hops from the source to the destination) will be saved in the header of the data packets. The intermediate nodes just read the packet header and forward the packet to the next hop. Furthermore, to overcome some drawbacks of the source routing, the route recovery is introduced.

### A. Topology Sensing

The topology sensing is to make the nodes get to know the topology information of the network, which includes link sensing, neighbor detection and topology discovery. This part gets benefit from MPRs as well as OLSR to minimize the flooding of broadcast packets in the network by reducing duplicate retransmissions in the same region.

By sending the routing control messages proactively, the node could get to know the topology of the network: its neighbors, 2-hop neighbors and other links, which are saved in neighbor set, 2-hop neighbors set and topology set respectively.

### B. Routes Computation

Contrary to classical OLSR, routes are not determined by nodes each time they receive a new Topology Control message, but only if need be, in order to avoid the loud computation of several routes for every possible destination. When a given source must send packets, the route computation procedure uses the algorithm shown in Figure 1.

The general principle of this algorithm is at step $i$ to look for the shortest path $P_i$ to the destination $d$. Then the edges in $P_i$ or pointing to $P_i$ have their cost increased in order to prevent the next steps to use similar path. $f_p$ is used to increase costs of arcs belonging to the previously path $P_i$ (or which opposite arcs belong to it). This encourages future paths to use different arcs but not different vertices. $f_e$ is used to increase costs of the arcs who lead to vertices of the previous path $P_i$. We can choose different $f_p$ and $f_e$ to get link-disjoint path or node-disjoint routes as necessary.

```
MultiPathDijkstra(s,d,G,N)
    c_1 ← c
    G_1 ← G
    for i ← 1 to N do
        SourceTree_i ← Dijkstra(G_i, s)
        P_i ← GetPath(SourceTree_i, d)
        if e or Reverse(e) is in P_i then
            c_{i+1}(e) ← f_p(c_i(e))
        else if Head(e) is in P_i then
            c_{i+1}(e) ← f_e(c_i(e))
        else
            c_{i+1}(e) ← c_i(e)
        end if
    end for
```

**Figure 1 The Multipath Dijkstra Algorithm**

In Figure 1, *Dijkstra(G,n)* is the standard Dijkstra's algorithm which provides the source tree of shortest paths from vertex *n* in graph *G*; *GetPath(ST,d)* is the function that extracts the shortest-path to *n* from the source tree *ST*; *Reverse(e)* gives the opposite edge of *e* ; *Head(e)* provides the vertex edge *e* points to.

Another possible solution is instead of increasing the cost of the arc, we delete the related node from the node set so that we can get totally node-disjoint routes, which is the ideal case [8]. However, in the real scenario, especially the cases that the nodes are sparse, the delete of some "key" nodes might prevent from finding new route. As shown in Figure 2, *A* is trying to send a packet to *E*. Initially, the costs of all the links are set to 1. It first find route A→B→C→D→E according to the Dijkstra algorithm. And to get the second route, it will increase the costs of used links, according to $f_p$, which are A→B, B→C, C→D and D→E in this case and also the links that lead to these nodes according to $f_e$ (not shown in the figure) . Then we use the Dijkstra algorithm again to get A→F→G→D→E. If we delete node *B, C* and *D* after we find the first route, there is no way to find the second one.

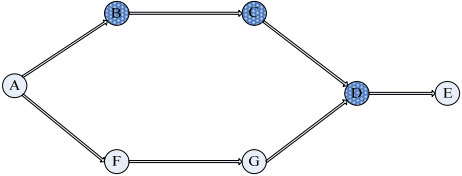

**Figure 2 Multiple Dijkstra Algorithm in sparse case**

Figure 3 gives the simulation results of the algorithm in the scenario of 300 nodes.

As shown in the figure, by using the Multipath Dijkstra Algorithm, we can get three node-disjoint routes. When the number of routes required comes to ten, we get ten optimal routes (some nodes might be shared when node-disjoint routes are unavailable) according to the cost functions. In this article, only three or four routes are used.

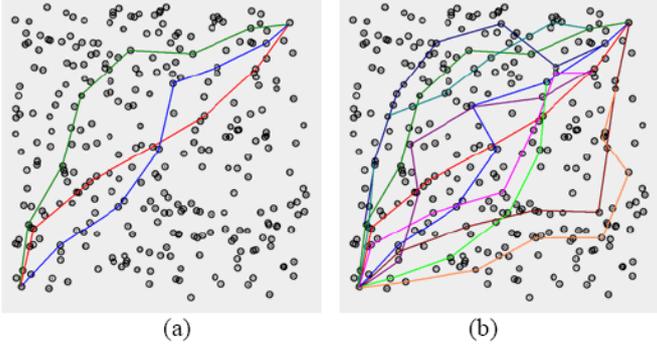

**Figure 3 Multiple Dijkstra Algorithm with $f_p(c) = f_e(c) = 2c$**
**(a) Three routes (b) Ten routes**

### C. Route Recovery

In the classical OLSR, the hop-by-hop routing is used, which means when a packet reaches an intermediate node, the protocol will check the routing table of the local node and then forward the packet to the next hop.

In contrast, in MP-OLSR, we use the semi-source routing approach. It will help the source node keep good control of the packets which will be forwarded in the multipath. However, in the mean time, the pure source routing might cause two problems: Firstly, the information in the source node might be not new enough because it needs time to flood the topology control messages to the whole network. It means when computing the routes, the source node might use the links that does not exist anymore. Secondly, even when the information in the source node is updated, the topology might change during the forwarding of the packet. Both of them will cause the failure of the packets forwarding.

To solve these problems, the *route recovery* is used: before a medium node trying to forward the packet, the node first check if the next hop in the source route of the packet is one of its neighbors. If yes, the packet is forwarded as it should be. If no, then it's possible that the "next hop" has moved out of the transmission range of the node. Then it is necessary to recompute the route and forward the packet through the new route.

For example, as show in Figure 4, node *A* is trying to forward packets to *D*. The original multiple source routes are A→B→C→D and A→E→F→G→D. However, node *G* moves out the transmission range of *F* and makes the second route unavailable anymore. When a packet is being forwarded through the second route, according to the source route stored in the packet header, and get to node *F*, it will first check if it's next hop *G* in node *F*'s neighbor set or not. If not, node *F* will recompute the route to *D*, and get *F→H→D*, then forward the packet through the new route.

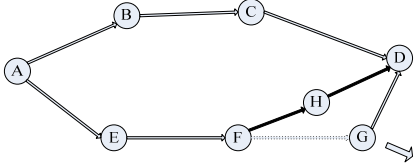

**Figure 4 Route Recovery**

Because the *Route Recovery* does not need to check the network information but just the topology information saved in the local node, it will not cause much extra delay and effectively improve the delivery ratio.

## IV. THE MULTIPLE DESCRIPTION CODING

By adding redundancy to information streams and splitting them up into several sub-streams, we can improve the integrity of data, especially by sending these sub-streams along different paths from the source to the destination. This kind of transformation is called Multiple Description Coding (see [9] for a detailed review).

Given a piece of information *I*, a multiple description coding method generates *N* independently communicable packets ($D_1, D_2,..., D_N$). Each description $D_i$ is generally much smaller than the original information. In the MDC scheme, the more descriptions are received, the closer to *I* is the reconstructed information *Î*. In our case, we just assume that it exists an integer *M* ($0<M≤N$) such that every subset of descriptions containing at least *M* different descriptions is sufficient to rebuild entirely *I*. Thus, the higher is *M*, the lower is the redundancy. In particular, *M* = *1* (respectively *M* =*N*) corresponds to the case where $(D_i)_{i\in[1,N]}$ are copies of *I* (respectively where $(D_i)_{i\in[1,N]}$ are different pieces of *I*).

In MP-OSLR, the Mojette transform [10] is used to produce different projections of the original information, each one being sent along a specific path. This discrete form of the Radon transform only requires the addition operation and is exactly invertible. For a detailed review, please refer to [11].

However, applying MDC coding to original packets may significantly increase the total number of packets that are transmitted in the network, which may results in new congestion.

A possible solution consists in setting up a sending buffer and then performing the MDC on its content (that corresponds to a group of original packets).The procedure is shown in Figure 5.

## V. SIMULATION AND PERFORMANCE ANALYSIS

### A. Environment and Assumption

The proposed algorithm is simulated on NS2. The channel capacity of mobile hosts was set to 11Mbps. A two-ray ground reflection model, which considers both the direct and a ground reflection path, was used as radio propagation model. We use the DCF (distributed coordination function) of IEEE 802.11 for wireless LAN as the MAC layer protocol. It has functionality to notify the network layer about link breakage.

In the simulations, there are 50 nodes move in a 1000m×1000m square region for 200 seconds simulation time. The random waypoint mobility model is used and there is no data packet transmission in the first 20 seconds to make sure the nodes are well distributed and there is sufficient time for the nodes to finish the initialization process. All the nodes have the same transmission range of 250 meters.

In each simulation, there are 30 CBR (Constant Bit Rate) sources, generating 10 packets per second with the size of 512 bytes.

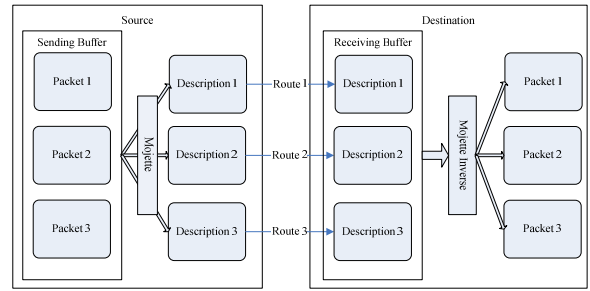

**Figure 5 The Multiple Description Coding**

### B. Performance metrics

The following metrics are used to evaluate the performance of the protocols:

*Packet delivery ratio*: the ratio of the data packets delivered to the destination.

*Routing load*: gives the number of routing packets over the number of received data packets. Each routing packet sent or forwarded by a mobile is counted.

*Average end-to-end delay*: The end-to-end delay is averaged over all surviving data packets form the sources to the destinations. It includes queuing delay and propagation delay.

*Load balancing*: We use a graph $G=(V,E)$ to denote the network, where V is the node set and E is the link set. We define a state function of $f: V→I$ where *I* is the set of positive integers. $f(v)$ represents the number of data packets forwarded at node *v*. Let coefficient of variation $CoV(f)$ = standard deviation of *f* / mean of *f*. We use $CoV(f)$ as a metric to evaluate the load balancing. The smaller the $CoV(f)$, the better the load balancing.

## C. Simulation results

We compared the performance of OLSR and MP-OLSR in the simulations.

Except for the difference of the protocols, two different kinds of strategy to discover the link failures are considered. The first one is the proactive way: when a node does not receive *HELLO* messages continuously from its former neighbors, it considers that the wireless link between them is broken and will delete the corresponding node from the neighbor set. The second strategy is to use link-layer feedback. When a node is trying to forward a packet to the next hop that has been out of its transmission range, its link layer will drop the packet because of *MAC_RET (*REtry Timeout). In the mean time, the link layer will give a feedback to the routing layer and inform the lost of the link. We also compared both of the strategies in the simulation.

The simulations are taken in 4 protocols:
- *The original OLSR*: It is the original single path routing protocol. It discovers the link failure in a proactive way. The implementation of UM-OLSR [12] is used here;
- *OLSR with link layer feedback* [12];
- *SR-MPOLSR:* MP-OLSR with link layer feedback, use source routing only. The incremental functions are: $f_p(c) = f_e(c) = 2c$;
- *RE-MPOLSR:* MP-OLSR with route recovery and link layer feedback. The incremental functions are: $f_p(c) = f_e(c) = 2c$.

Figure 6 shows the delivery ratio of the simulations. As we can see from the figure, the OLSR with feedback has better delivery ratio than the original OLSR. This is because the protocol with feedback could detect a link failure as soon as the first packet is dropped and recomputed the routing table. In the contrast, the original OLSR tends to continue to send the packets through a failed link until it finds out that it is not able to receive a *HELLO* message from the original neighbor. Both with the link layer feedback, the SR-MPOLSR's delivery ratio is about ten percent lower than the OLSR. This is because of the drawbacks of source routing that have been mentioned in the previous section and compared with OLSR, which always sends the packets through the best route, the SR-MPOLSR sends the packets in multiple routes, some of which might be more unreliable. However, the RE-MPOLSR gives better delivery ratio thanks to the route recovery.

Figure 7 gives the simulation results of routing load. Because in all the scenarios, the number of the generated control messages is the same, so the protocol with higher deliver ratio tends to have lower routing load.

Figure 8 shows the average end-to-end delay. It includes the queue delay in every node and the propagation delay from the source to the destination. The multipath routing could reduce the queue delay because the traffic is distributed in different routes. On the other hand, it might increase the propagation delay because some of the packets are sent through the sub-optimal route. As we can see from the figure, compared with the single path protocols, although the multipath routing might result in more propagation delay, it could effectively reduce the queue delay, and tends to have lower end-to-end delay. The figure also shows that although RE-MPOLSR might spend more time in recomputing and recovering the path, it still has the lowest delay because the recovery mechanism could avoid sending data packets through a failed route, which will result in more retry time and the network congestion. The multipath protocols also provide more stable end-to-end delay.

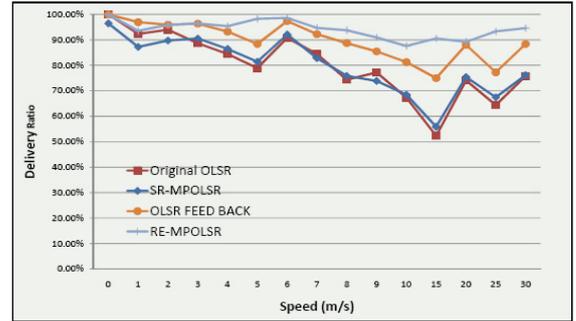
**Figure 6 Delivery Ratio**

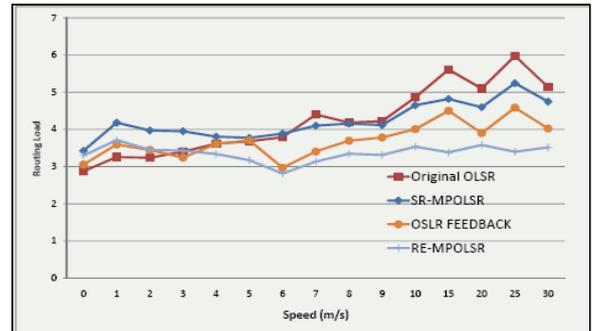
**Figure 7 Routing Load**

Figure 9 demonstrates the results of load balancing. The CoV of network load for the unipath routing is higher than the multipath routing because the packets are distributed along the different routes in multipath routing. Also in the scenario with high node mobility, the network tends to have better load balancing because more nodes are included in the transmission in a mobile networks.

The simulation of MP-OLSR with MDC is also taken. The gain of delivery ratio is very little (one or two percent) compared with the RE-MPOLSR in the scenario of 50 nodes. This is because given the low density of the nodes, sometimes the multiple paths will share a lot of links or even there is just no route at all. Then it is hard to make benefits from the MDC. In fact, the MDC tends to be more suitable for large and dense networks.

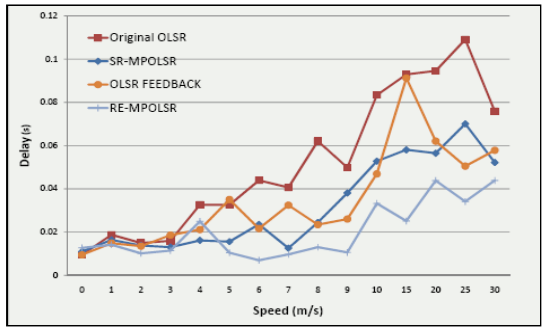

**Figure 8 End to end delay**

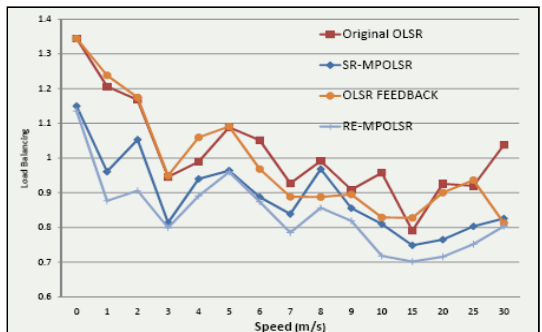

**Figure 9 Load Balancing**

Figure 10 gives the delivery ratio in the scenario of 100 nodes. This figure compares 3 protocols: RE-MPOLSR, OLSR Feedback and MDC-OLSR in which the number of descriptions is 4 (i.e. number of routes is 4) and the number of useful descriptions in the reconstruction of initial information is 2. So with higher density of the nodes, MP-OLSR with MDC will have more reliable data transmission by offering better delivery ratio. It is about 10% for speed range between 6m/s and 10m/s.

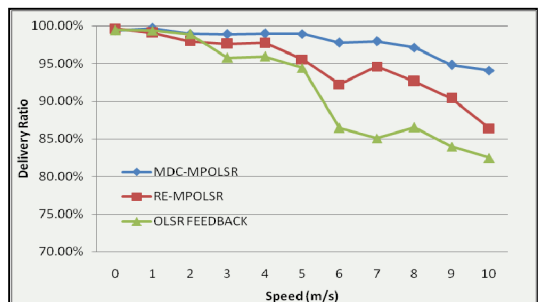

**Figure 10 Delivery Ratio in the scenario of 100 nodes**

## VI. CONCLUSIONS AND FUTURE WORKS

In this paper, we discussed a multipath extension to OLSR. Our Multipath Dijkstra Algorithm is used to discover the disjoint routes and the source routing with route recovery is performed to forward the packets. The new multipath protocol could offer better end-to-end delay and load balancing. Because of the overlapping radio-coverage of neighbor nodes and that the limitation of the MAC protocol can result in strong interdependence between multi-routes, the multipath protocol performance gains achieved in ad hoc networks is not as much as in the wired Internet. Moreover, a pure source routing strategy, can appear as a sub-utilization of a good topology knowledge inherent in proactive behavior. We observe that the SR-MPOLSR has worse packet delivery ratio than the OLSR with link layer feedback. However, with the routing recovery, the MP-OLSR can achieve the best performance. To meet the requirement for a reliable transmission, the multiple routes are exploited by a multiple description coding based on Mojette Transform.

The future research includes refining the incremental functions of $f_p$ and $f_e$ to make them adaptive to the specific network and optimize the redundancy allocation for MDC in the data transmission. More largely, more precise simulations of physical layer within NS2 are engaged. Model of radio propagation [13] in the middle of buildings could validate the importance of the multipath routing for ad hoc networks.